\documentclass{article}

\usepackage{arxiv}

\usepackage[utf8]{inputenc}
\usepackage[T1]{fontenc}    
\usepackage{hyperref}       
\usepackage{url}            
\usepackage{booktabs}       
\usepackage{amsfonts}       
\usepackage{amsmath}
\usepackage{nicefrac}       
\usepackage{microtype}                   
\usepackage{graphicx}
\usepackage{verbatim}

\usepackage{amsmath,amssymb,amsthm}

\newtheorem{lemma}{Lemma}

\usepackage{tikz}
\usetikzlibrary{calc}
\usetikzlibrary{shapes.geometric, positioning, arrows.meta}

\usepackage{multirow}

\usepackage{enumitem}
\setlist[itemize]{labelindent=1em,leftmargin=2em,labelsep=0.5em}

\usepackage{braket}

\makeatletter
\def\blfootnote{\xdef\@thefnmark{}\@footnotetext}
\makeatother

\usepackage{fancyhdr}
\begin{document}

\title{Full characterization of informative subsets\\ in Quantum Encrypted Cloning}

\author{
  \href{https://orcid.org/0000-0001-5186-0199}{\includegraphics[scale=0.06]{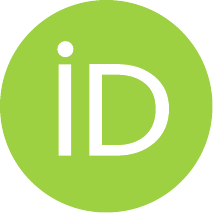}\hspace{1mm} Gabriele Gianini}\\
  Universit\`a degli Studi di Milano-Bicocca \\
  Milan, Italy \\
  \texttt{gabriele.gianini@unimib.it}\\ \vspace{1pt}
\and
  \href{https://orcid.org/0000-0003-1737-6218}{\includegraphics[scale=0.06]{orcid.pdf}\hspace{1mm}
  Stelvio Cimato}\\
  Universit\`a degli Studi di Milano\\
  Milan, Italy\\
  \texttt{stelvio.cimato@unimi.it}\\ \vspace{1pt}
\and
  \href{https://orcid.org/0000-0002-3299-448X}{\includegraphics[scale=0.06]{orcid.pdf}\hspace{1mm}
  Jianyi Lin}\\
  Universit\`a Cattolica del Sacro Cuore\\
  Milan, Italy\\
  \texttt{jianyi.lin@unicatt.it}\\ \vspace{1pt}
\and
\href{https://orcid.org/0000-0002-9585-7810}{\includegraphics[scale=0.06]{orcid.pdf}\hspace{1mm}
  Omar Hasan}\\
  Institut National des Sciences Appliqu\'ees de Lyon\\
  Lyon, France \\
  \texttt{omar.hasan@liris.cnrs.fr}\\ \vspace{1pt}
\and
  \href{https://orcid.org/0000-0002-9557-6496}{\includegraphics[scale=0.06]{orcid.pdf}\hspace{1mm}
  Ernesto Damiani}\\
  Universit\`a degli Studi di Milano\\
  Milan, Italy\\
  \texttt{ernesto.damiani@unimi.it}
}
\date{}
\maketitle
\begin{abstract}
Quantum encrypted cloning, introduced by Yamaguchi and Kempf, is a Pauli-based protocol that distributes an unknown input qubit into multiple encrypted signal-noise pairs in such a way that redundancy is created without violating the no-cloning theorem, since at most one clone can later be perfectly recovered through an appropriate decoding procedure. In previous work we showed that unauthorized subsets of the storage register are not, in general, completely uninformative, and we identified a parity-dependent leakage pattern. In the present work we extend the analysis to subsets that also include the transformed source qubit \(A\). Exploiting the purity of the global encoded state and the complementarity between storage-only subsets and subsets containing \(A\), we derive a full classification of the informativeness of all sets of the form \(H=\{A\}\cup C\). We show that these subsets are fully informative in the generic case. Two exceptions arise. First, if all pairs are incomplete and \(|C|<n\), then the reduced state is completely uninformative. Second, if \(|C|=n\), \(n\) is odd, and the number \(q\) of signal qubits in \(C\) is even, then the reduced state is partially informative. In this latter case, the residual dependence on the input state is confined to the \(y\)-component of the Bloch vector. These results provide a complete parity-based characterization of leakage for subsets containing the transformed input qubit.
\keywords{
Encrypted cloning;
Quantum redundancy;
Information leakage;
Authorized subsets;
Confidentiality;
}
\end{abstract}

\section{Introduction}\label{sec:intro}

Yamaguchi and Kempf \cite{yamaguchi2026encrypted,yamaguchi2026experimental} introduced \textit{encrypted cloning}, a Pauli-operator-based protocol that enables the creation of multiple encrypted clones of an unknown quantum state without violating the no-cloning theorem, since at most one clone can later be recovered through key-consuming decryption. The protocol was later extended by Cear\u{a} to arbitrary finite dimensions \cite{ceara2026cloningencryptedquantumstates}.

In encrypted cloning, redundancy is not introduced as a conventional secrecy primitive, but as a way of distributing recoverable quantum information across several signal-noise subsystems while preserving consistency with the no-cloning theorem. For this reason, confidentiality is not guaranteed by design and must be analyzed explicitly. In particular, the relevant question is not only which subsets are authorized for perfect recovery, but also which non-authorized subsets may nevertheless retain some dependence on the input state.

In a previous work \cite{gianini2026encrypted} we investigated the informativeness properties of subsets of the storage register, namely subsets formed only by signal and noise qubits and not containing the transformed input qubit \(A\). We showed that unauthorized subsets are not necessarily completely uninformative: besides fully informative and completely uninformative sets, there exists an intermediate class of partially informative subsets. More precisely, for subsets of size \(n\) containing exactly one representative from each signal-noise pair, the reduced state is completely uninformative when \(n\) is even, whereas for odd \(n\) a residual dependence on the input state survives exactly when the number of signal qubits in the subset is odd. In that case, the leakage is confined to the \(y\)-component of the Bloch vector. These results were later extended by Bai, Zhou, and Luo \cite{bai2026classification} to the qudit version of the protocol, where the presence or absence of leakage is determined by the solution set of a system of congruences depending on the dimension \(d\) and on the composition of the subset.

In the present work we complete our original analysis ($d=2$) by considering subsets that also include the transformed source qubit \(A\). This extension is not merely a change of subsystem. Since the encoded state on \(A\cup{\mathcal R}_n\) is pure, the informativeness of subsets containing \(A\) is tightly related to that of complementary storage-only subsets. This allows us to transfer the previous classification on \({\mathcal R}_n\) into a corresponding classification for sets of the form
\[
H=\{A\}\cup C,
\qquad
C\subseteq{\mathcal R}_n,
\]
and to derive a three-way partition into fully informative, completely uninformative, and partially informative subsets.

The resulting structure turns out to be strongly constrained. Subsets containing \(A\) are generically more informative than storage-only subsets, but two exceptional regimes remain. When all signal-noise pairs are incomplete and \(|C|<n\), the reduced state is completely uninformative. When \(|C|=n\), \(n\) is odd, and the number \(q\) of signal qubits in \(C\) is even, the reduced state is partially informative. In this latter case, the leakage again survives only through the \(y\)-Bloch channel. Thus, the extension to subsets containing \(A\) does not destroy the parity-based structure found for storage-only subsets, but reshapes it in a complementary way.

From a broader perspective, these results show that encrypted cloning can be viewed as a pure-state quantum encoding with a nontrivial access structure: some subsets allow perfect reconstruction, some are completely decoupled from the input state, and others retain only partial information. In this sense, the protocol exhibits features reminiscent of quantum secret sharing, although not of a perfect one, because intermediate subsets may still leak restricted information about the secret state.

The paper is organized as follows. In Section~\ref{sec:info} we recall the Pauli-based encrypted-cloning protocol and the previous classification of storage-only subsets. In Section~\ref{sec:including-A} we extend the analysis to subsets containing \(A\), derive their informativeness properties by complementarity, and obtain the explicit reduced-state expressions for the case \(|C|=n\), identifying the corresponding leakage channel. Section~\ref{sec:conclusions} concludes the paper.

\section{Formal background}

We recall the encrypted-cloning protocol introduced in \cite{yamaguchi2026encrypted}, and the leakage results presented in \cite{gianini2026encrypted}.

\subsection{The Pauli-based encoding}
Let \(A\) denote the input qubit, prepared in an unknown pure state $\ket{\psi}_A$.
To generate \(n>1\) encrypted clones, the protocol introduces \(n\) Bell pairs
\[
\ket{\phi}_{S_iN_i}=\frac{1}{\sqrt 2}(\ket{00}+\ket{11}),
\qquad i=1,\dots,n,
\]
where \(S_i\) is the \(i\)-th signal qubit and \(N_i\) is the corresponding noise, or key, qubit. 

The encrypted-cloning encoding acts jointly on the input qubit \(A\) and on the signal qubits \(S_1,\dots,S_n\), while leaving the noise qubits \(N_1,\dots,N_n\) unaffected. To encode a pure state $\ket{\psi}$ into $n$ encrypted clones, Yamaguchi and Kempf use the following unitary,  
based on the Pauli operators:
\(
\sigma_0=I,\,
\sigma_1=X,\,
\sigma_2=Y,\,
\sigma_3=Z.
\)
\[
U_{\mathrm{enc}}^{(n)}
=
\frac{1}{2}\sum_{\mu=0}^3 \alpha_\mu^{-1}\sigma_\mu^{(A)}
\otimes\left(\bigotimes_{i=1}^n \sigma_\mu^{(S_i)}\right),
\]
with
\begin{equation}\label{eq:alphas}
\alpha_0=1,\qquad
\alpha_1=\alpha_3=i,\qquad
\alpha_2=-i^{\,n+1}.
\end{equation}

The resulting encoded state (see \cite{yamaguchi2026encrypted}, Eq.~(7)) is
\[
\ket{\Psi_{\mathrm{enc}}}
=
U_{\mathrm{enc}}^{(n)}
\left[
\ket{\psi}_A \otimes
\left(\bigotimes_{i=1}^n \ket{\phi}_{S_iN_i}\right)
\right]
=
\frac{1}{2}\sum_{\mu=0}^3 \alpha_\mu^{-1}\sigma_\mu^{(A)}\ket{\psi}_A
\otimes
\left(\bigotimes_{i=1}^n \ket{\phi_\mu}_{S_iN_i}\right),
\]
where we use the shorthand
\[
\ket{\phi_\mu}_{S_iN_i}
\equiv
\left(\sigma_\mu^{(S_i)}\otimes I^{(N_i)}\right)\ket{\phi}_{S_iN_i}.
\]

The corresponding density matrix is
%
%
%
%
\begin{equation}\label{eq:encoded-state}
\rho_{\mathrm{enc}}^{(n)}
=
\frac{1}{4}
\sum_{\mu,\nu=0}^3
\alpha_\mu^{-1}\alpha_\nu\,
\left(\sigma_\mu^{(A)} \ket{\psi}\!\bra{\psi}\,\sigma_\nu^{(A)}\right)
\otimes
\left(
\bigotimes_{i=1}^n
\ket{\phi_\mu}\!\bra{\phi_\nu}_{\,S_iN_i}
\right).
\end{equation}

This decomposition shows that encrypted cloning is governed by the coherent superposition of four Pauli branches, which gives rise to sixteen terms in the density-matrix expansion. 

\subsection{Subsets, Authorization and Informativeness}\label{sec:info}

Consider the \textit{storage register}, i.e. the subset  of the encoded state not including the qubit $A$, consisting of $2n$ qubits: 
$$
{\mathcal R}_n\equiv \{S_1,N_1,S_2,N_2,\dots,S_n,N_n\}
$$

Yamaguchi and Kempf \cite{yamaguchi2026encrypted} showed which subsets of the storage register are authorized and which are unauthorized. Authorized sets are fully informative, i.e. hold enough information to enable the full recovery of the original state. 

Unauthorized access does not by itself imply complete absence of informativeness about the input state. In the work \cite{gianini2026encrypted} we investigated all the unauthorized sets of the register and characterized their informativeness.
Hereafter we recall the authorization rules from \cite{yamaguchi2026encrypted} and the informativeness rules from \cite{gianini2026encrypted}. In the next section we extend the analysis to the subsets of the encoded state given by the union of $A$ with subset of the storage register.

Let us define some important conditions for a set $B$. We explicitly state also the relevant complementary conditions for the sake of presentation clarity.

An important aspect concerns the presence of at least a complete signal-noise pair:
\begin{itemize}
\item {\bf [FULL-PAIR]} condition. $B$  contains at least one complete pair: $\exists\,k:(S_k\in B)\wedge  (N_k \in B)$; 
\item \textbf{[ALL-PAIRS-INCOMPLETE]} condition. No full pair is present:\ all the pairs are either missing or incomplete: $\forall\,j: (S_j\notin B)\vee  (N_j \notin B)$.
It is the complementary of the previous one.\end{itemize}
Another aspect refers to the ensemble of pairs and concerns the presence of at least a representative for each signal-noise pair, either a signal qubit or a noise qubit.
\begin{itemize}
\item {\bf [SPAN]}, or all-pairs-span, condition. $B$ contains at least one element of each pair: $\forall\,j: (S_j\in B)\vee  (N_j \in B)$. 
\item {\bf [MISSING-PAIR]}, or missing-pair condition. Not all the pairs have a representative, since at least one pair is completely missing: $\exists\,k: (S_k\notin B)\wedge  (N_k \notin B)$.
This is the complementary of the previous one.
\end{itemize}

Notice for future reference that given the set $C$ complementary of $B$ within ${\mathcal R}_n$, so that $(S_j\in B) \iff (S_j \notin C)$, and similarly for $N_j$, then the following implications between conditions hold
\begin{itemize}
\item
if either $B$ or $C$ fulfills the FULL-PAIR condition, then the other fulfills the MISSING-PAIR condition, and vice versa;
\item
if one fulfills the SPAN condition, the other fulfills the ALL-PAIRS-INCOMPLETE condition, and vice versa.
\end{itemize}

Moreover, notice that $|B|>n$ implies that $B$ fulfills the FULL-PAIR condition, and that $|B|<n$ implies that $B$ fulfills the MISSING-PAIR condition. Both follow by a pigeonhole argument: a subset of more than $n$ qubits meeting every pair must contain both members of at least one pair, whereas a subset of fewer than $n$ qubits cannot meet all $n$ pairs. In particular, the ALL-PAIRS-INCOMPLETE condition implies $|B|\leq n$.

\paragraph{Authorization.} The authorized, i.e. fully informative, sets are defined in \cite{yamaguchi2026encrypted} by the following rules: 
\begin{itemize}
\item A subset $B\subseteq {\mathcal R}_n$ is authorized if fulfills the SPAN  condition and the {{FULL-PAIR}} condition. E.g.,
\begin{itemize}
\item
the set containing one encrypted clone qubit and all the noise qubits is authorized;
\end{itemize}
A subset $B\subseteq {\mathcal R}_n$ is unauthorized whenever it fails to satisfy either of those two conditions. E.g.,
\begin{itemize}
\item
any subsystem consisting of \((n-1)\) complete pairs is unauthorized
\end{itemize}
\item
Any subsystem consisting of all \(n\) noise qubits is unauthorized.
\end{itemize}
\paragraph{Informativeness.}
If a subset $B$ is authorized we say it is \textit{fully informative}. Given an unauthorized subset \(B\), we call it \textit{completely uninformative} if its reduced state
\(
\rho_B(\psi)
\)
is independent of the input state \(\ket{\psi}\). In contrast, we say that an unauthorized subset \(B\) is \textit{partially informative} if \(\rho_B(\psi)\) retains a nontrivial dependence on \(\ket{\psi}\).
A special case of \textit{completely uninformative} state is a maximally mixed state.

In the work \cite{gianini2026encrypted} we characterized the informativeness of the subsets of the storage register as follows. Let $B\subseteq {\mathcal R}_n$.
\begin{itemize}
\item If $B$ fulfills the MISSING-PAIR condition, it is \textit{completely uninformative}.
\begin{itemize}
\item if $|B|<n$, this necessarily happens.
\end{itemize}
\item If $B$ fulfills the SPAN condition, it must have at least size $n$, then
\begin{itemize}
\item when $|B| > n$ the set $B$ must also fulfill the FULL-PAIR condition and is \textit{fully informative} 
\item when $|B|=n$, 
\begin{itemize}
\item if $n$ is even, $B$ is \textit{completely uninformative};
\item if $n$ is odd, 
\begin{itemize}
\item if $p$ is even $B$ is \textit{completely uninformative};
\item if $p$ is odd $B$ is \textit{partially informative};
\end{itemize}
\end{itemize}
\end{itemize}
\end{itemize}
The reduced state for sets of size $|B|=n$ for the case $B\subseteq {\mathcal R}_n$ completely uninformative is the maximally mixed state $I^{\otimes n}\big/2^{n}$, whereas for the partially informative case ($n$ odd and $p$ odd) it is the following
\begin{equation}
\rho_{S_1,\ldots,S_p,N_{p+1},\ldots,N_n}
=
\frac{1}{2^n}
\left(I^{\otimes n}+(-1)^{{(n-1)}/{2}}\,y\,Y^{\otimes n}\right),
\end{equation}

\begin{figure}[t]
\centering
\begin{tikzpicture}[
  >=Latex,
  font=\small,
  level distance=11mm,
  xlevel1/.style={sibling distance=95mm},
  xlevel2/.style={sibling distance=72mm},
  xlevel3/.style={sibling distance=43mm},
  xlevel4/.style={sibling distance=32mm},
  xlevel5/.style={sibling distance=18mm},
  root/.style={
    rectangle,
    rounded corners,    
    draw,
    align=center,
    fill=white,
    minimum height=8mm,
    text width=21mm,
    inner sep=0pt
  },
  internal/.style={
    rectangle,
    draw,
    fill=white,
    align=center,
    inner sep=3pt,
    minimum height=6mm
  },
  leafbase/.style={
    rectangle,
    rounded corners,
    draw,
    align=center,
    inner sep=4pt,
    minimum height=8mm,
    text width=21mm
  },
  leafU/.style={
    leafbase,
    fill=red!12,
    draw=red!50!black
  },
  leafF/.style={
    leafbase,
    fill=green!12,
    draw=green!45!black
  },
  leafP/.style={
    leafbase,
    fill=yellow!18,
    draw=yellow!50!black
  },
  dummy/.style={
    draw=none,
    fill=none,
    inner sep=0pt,
    minimum size=0pt
  },
  ghost edge/.style={
    edge from parent/.style={draw, -}
  },
  leaf edge/.style={
    edge from parent/.style={draw, -}
  },
  edge from parent/.style={draw, -Latex}
]

\node[root] {$B\subseteq {\mathcal R}_n$}
child[xlevel1] {
  node[internal] {MISSING-PAIR}
  child[ghost edge] {
    node[dummy] {}
    child[ghost edge] {
      node[dummy] {}
      child[ghost edge] {
        node[dummy] {}
        child[leaf edge] {
          node[leafU] {$B$ \textit{completely\\ uninformative}}
        }
      }
    }
  }
}
child[xlevel1] {
  node[internal] {SPAN}
  child[xlevel2] {
    node[internal] {$|B|=n$}
    child[xlevel3] {
      node[internal] {$n$ even}
      child[ghost edge] {
        node[dummy] {}
        child[leaf edge] {
          node[leafU] {$B$ \textit{completely\\ uninformative}}
        }
      }
    }
    child[xlevel3] {
      node[internal] {$n$ odd}
      child[xlevel4] {
        node[internal] {$p$ even}
        child[leaf edge] {
          node[leafU] {$B$ \textit{completely\\ uninformative}}
        }
      }
      child[xlevel4] {
        node[internal] {$p$ odd}
        child[leaf edge] {
          node[leafP] {$B$ \textit{partially\\ informative}}
        }
      }
    }
  }
  child[xlevel2] {
    node[internal] {$|B|>n$}
    child[xlevel3] {
      node[internal] {FULL-PAIR}
      child[ghost edge] {
        node[dummy] {}
        child[leaf edge] {
          node[leafF] {$B$ \textit{fully\\ informative}}
        }
      }
    }
  }
};

\end{tikzpicture}
\caption{Schema of the informativeness criteria for $B\subseteq {\mathcal R}_n$, with $n$ total number of signal qubits, and $p\leq n$ number of signal qubits in $B$.}
\end{figure}
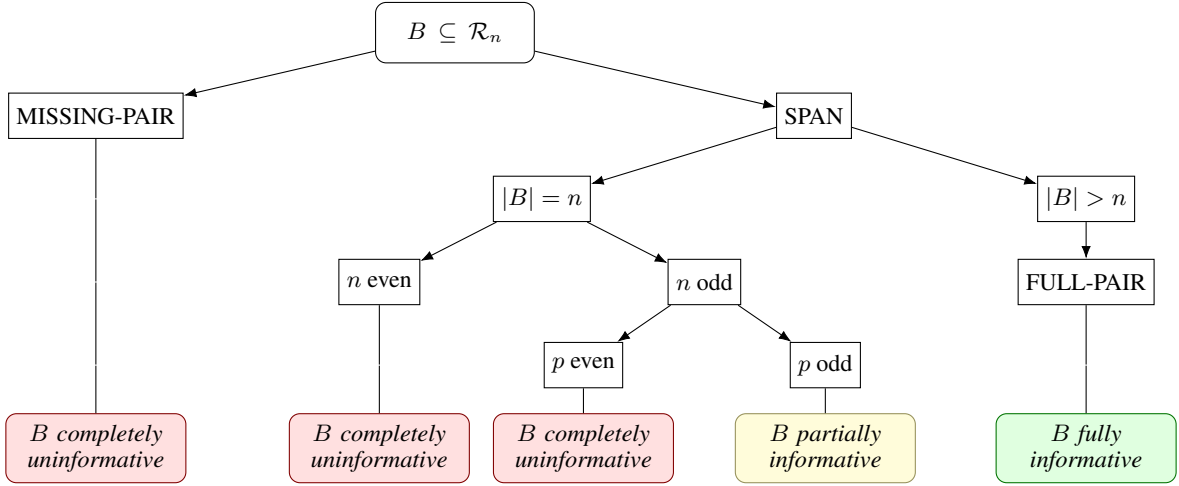

\newpage
\section{Including the source qubit \(A\)}\label{sec:including-A}

In this section, we extend the leakage analysis beyond the encrypted-clone storage register by considering subsets that also include the source qubit \(A\). This extension leads to a related parity dependent structure. 

Hereafter, first we outline the approach, then derive the informativeness properties of the sets that include the qubit $A$, and finally work out the reduced form expression for the sets where $|C|=n$, with the aim of obtaining the form of partially informative sets and identifying the leakage channel: it will turn out to be the $y$-Bloch channel.

\subsection{Outline of the approach}
To analyze subsets containing \(A\), we use the fact that the global state on
\(
A\,\cup{\mathcal R}_n
\)
remains pure.
\emph{Global purity allows us to characterize subsets containing \(A\) through their complementary subsystems}, whose informative properties were described in Section \ref{sec:info}. 

Specifically, for a subset \(H\subseteq A\cup{\mathcal R}_n\) with \(A\subseteq H\), one can relate the information accessible from \(H\) to the structure of its complement in the pure system on \(A\,\cup{\mathcal R}_n\). This turns the classification on \({\mathcal R}_n\), from the previous section, into a corresponding classification for subsets containing \(A\).

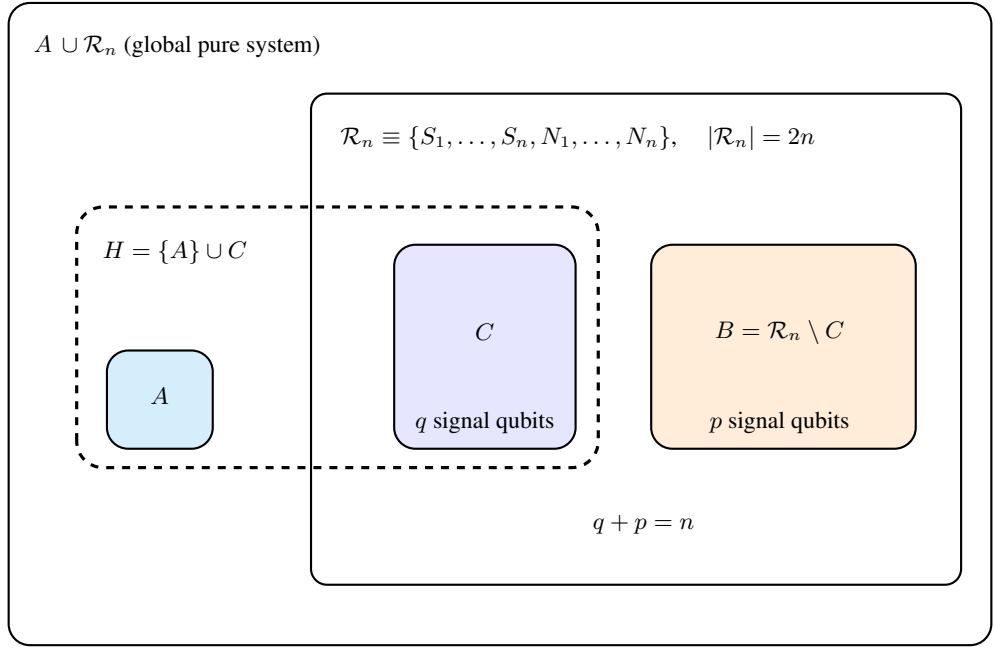
\begin{figure}[t]
\centering
\begin{tikzpicture}[
    font=\small,
    thick,
    every node/.style={align=center}
]

\draw[rounded corners=8pt] (0,0) rectangle (13,8.5);
\node[anchor=north west] at (0.2,8.2) {$A\,\cup {\mathcal R}_n$ (global pure system)};

\draw[rounded corners=6pt] (4,0.8) rectangle (12.6,7.3);
\node[anchor=north east] at (10.8,7.0) {${\mathcal R}_n\equiv\{S_1,\dots,S_n,N_1,\dots,N_n\}$, \quad$|{\mathcal R}_n|=2n$};

\draw[rounded corners=8pt, fill=blue!10] (5.1,2.6) rectangle (7.5,5.3);
\node at (6.3,4.15) {$C$};
\node at (6.3,2.95) {$q$ signal qubits};

\draw[rounded corners=8pt, fill=orange!15] (8.5,2.6) rectangle (12.0,5.3);
\node at (10.2,4.15) {$B={\mathcal R}_n\setminus C$};
\node at (10.2,2.95) {$p$ signal qubits};

\node at (8.4,1.60) {$q+p=n$};

\draw[rounded corners=8pt, fill=cyan!15] (1.3,2.6) rectangle (2.7,3.9);
\node at (2.0,3.3) {$A$};


\draw[dashed, very thick, rounded corners=10pt] (0.9,2.35) rectangle (7.8,5.8);
\node at (2.2,5.2) {$H=\{A\}\cup C$};

\end{tikzpicture}
\caption{Diagrammatic representation of a subset \(H=\{A\}\cup C\) and of the complementary storage-only subset \(B\).}
\label{fig:withA-complementarity-venn}
\end{figure}

The reasoning schema proceeds as follows (see Figure \ref{fig:withA-complementarity-venn}): we characterize the set $H$, containing $A$, through the union of $A$ with a subset $C$ of the storage register of $2n$ qubits ${\mathcal R}_n$; then we look at the set $B$ defined as the complementary of $C$ within ${\mathcal R}_n$; finally from the previously stated informativeness properties of $B$, we deduce the property of $H$. 

The formal basis of this reasoning is the following standard complementarity between recoverability and privacy, which we state in the form in which it will be used.

\begin{lemma}[Complementarity of recoverability and privacy]\label{lem:decoupling}
Let ${V:{\mathcal H}_L\to{\mathcal H}_H\otimes{\mathcal H}_B}$ be an isometry and, for an input state $\rho$ on ${\mathcal H}_L$, let
\[
\rho_H(\rho)=\operatorname{Tr}_B\big(V\rho V^{\dagger}\big),
\qquad
\rho_B(\rho)=\operatorname{Tr}_H\big(V\rho V^{\dagger}\big)
\]
be the reduced states on the two complementary subsystems. Then the following are equivalent:
\begin{itemize}
\item[(i)] there exists a completely positive trace-preserving map ${\mathcal D}$ acting on ${\mathcal H}_H$ alone such that ${\mathcal D}\big(\rho_H(\rho)\big)=\rho$ for every input state $\rho$;
\item[(ii)] $\rho_B(\rho)$ is independent of $\rho$.
\end{itemize}
\end{lemma}

Lemma \ref{lem:decoupling} is the information-disturbance tradeoff of quantum information theory: the information about an unknown state that can be recovered from a subsystem is exactly the information that has not leaked into the complementary one \cite{schumacher1998privacy}; for the approximate and quantitative form of the same tradeoff see \cite{kretschmann2008continuity}, and for a textbook treatment \cite{nielsen2010quantum}. It is also the mechanism that underlies the access structure of quantum secret sharing \cite{cleve1999share}. In the terminology of Section~\ref{sec:info}, condition (i) states that the subsystem is authorized, hence fully informative, and condition (ii) states that its complement is completely uninformative.

Since the encoded state on $A\cup{\mathcal R}_n$ is pure, the lemma applies to the bipartition $H\,|\,B$ in both directions. The informativeness class of \(H\) is therefore determined by the informativeness class of \(B\). In particular:

\begin{itemize}
    \item if \(B\) is authorized, then it is \textit{fully informative}; applying Lemma~\ref{lem:decoupling} with $B$ in the role of the recovering subsystem, condition (i) holds for $B$, hence condition (ii) holds for its complement, and \(H\) must be \textit{completely uninformative};
    \item if \(B\) is \textit{completely uninformative}, then condition (ii) of Lemma~\ref{lem:decoupling} holds for $B$, so condition (i) holds for \(H\): all the information needed to recover the input state is localized in \(H\), which is therefore authorized, i.e. \textit{fully informative};
    \item if \(B\) is \textit{partially informative}, then \(H\) can be neither authorized nor completely non-informative, for the previous two statements, and is therefore \textit{partially informative} as well.
\end{itemize}

\subsection{Informativeness of the sets including the qubit $A$}\label{sec:withA-class}

Explicitly, let
\[
H=\{A\}\cup C,
\qquad
C\subseteq {\mathcal R}_n,
\qquad
B\equiv{\mathcal R}_n\setminus C.
\]
Let us also denote the number of signal qubits in $C$ by $q$, and the number of signal qubits in $B$ by $p$, so that $p+q=n$. Note that in \cite{gianini2026encrypted} the number of signal qubits of the subset under analysis was denoted by $p$; here that role is played by $q$, since the subset under analysis is $C$.


\begin{itemize}
\item
if $C$ fulfills the FULL-PAIR condition, then $B$ fulfills the MISSING-PAIR condition and is completely uninformative;\ thus $H$ must be \textit{fully informative}, i.e. authorized.
\begin{itemize}
\item If $|C| > n$, then  $|B|<n$, and this is necessarily the case.
\end{itemize}
\item 
if  $C$ fulfills ALL-PAIR-INCOMPLETE condition, then $B$ fulfills the SPAN condition
\begin{itemize}
\item If $|C| < n$, then  $|B|>n$: $C$ fulfills the MISSING-PAIR condition, and $B$ the FULL-PAIR condition, which together with the SPAN condition makes $B$ authorized;\ thus $H$ is \textit{completely uninformative}.
\item if $|C|=n$, then also $|B|=n$
\begin{itemize}
\item if $n$ is even, $B$ is {completely uninformative}; thus $H$ is \textit{fully informative};
\item if $n$ is odd, then, denoting by $q$ the number of signal qubits in $C$ and by $p$ the number of signal qubits in $B$, we have that
\begin{itemize}
\item if $q$ is odd, then $p$ is even and $B$ is {completely uninformative}; thus, $H$ must be \textit{fully informative};
\item if $q$ is even, then $p$ is odd and $B$ is {partially informative}; thus $H$ must be \textit{partially informative}.
\end{itemize}
\end{itemize}
\end{itemize}
\end{itemize}

\begin{figure}[bt]
\centering
\begin{tikzpicture}[
  >=Latex,
  font=\small,
  level distance=11mm,
  xlevel1/.style={sibling distance=95mm},
  xlevel2/.style={sibling distance=82mm},
  xlevel3/.style={sibling distance=48mm},
  xlevel4/.style={sibling distance=35mm},
  xlevel5/.style={sibling distance=18mm},
  root/.style={
    rectangle,
    rounded corners,    
    draw,
    align=center,
    fill=white,
    minimum height=8mm,
    text width=26mm,
    inner sep=0pt
  },
  internal/.style={
    rectangle,
    draw,
    fill=white,
    align=center,
    inner sep=3pt,
    minimum height=6mm
  },
  leafbase/.style={
    rectangle,
    rounded corners,
    draw,
    align=center,
    inner sep=4pt,
    minimum height=8mm,
    text width=21mm
  },
  leafU/.style={
    leafbase,
    fill=red!12,
    draw=red!50!black
  },
  leafF/.style={
    leafbase,
    fill=green!12,
    draw=green!45!black
  },
  leafP/.style={
    leafbase,
    fill=yellow!18,
    draw=yellow!50!black
  },
  dummy/.style={
    draw=none,
    fill=none,
    inner sep=0pt,
    minimum size=0pt
  },
  ghost edge/.style={
    edge from parent/.style={draw, -}
  },
  leaf edge/.style={
    edge from parent/.style={draw, -}
  },
  edge from parent/.style={draw, -Latex}
]

\node[root] {$C = \big({\mathcal R}_n \setminus B\big)$}
child[xlevel1] {
  node[internal] {$C$ FULL-PAIR $\Rightarrow$\\ $\Rightarrow B$ MISSING-PAIR }
  child[ghost edge] {
    node[dummy] {}
    child[ghost edge] {
      node[dummy] {}
      child[ghost edge] {
        node[dummy] {}
        child[leaf edge] {
          node[leafF] {$H$ \textit{fully\\ informative}}
        }
      }
    }
  }
}
child[xlevel1] {
  node[internal] {$C$ ALL-PAIRS-INCOMPLETE  $\Rightarrow B$ SPAN}
  child[xlevel2] {
    node[internal] {$|C|=n$ $\Rightarrow$ $|B|=n$}
    child[xlevel3] {
      node[internal] {$n$ even}
      child[ghost edge] {
        node[dummy] {}
        child[leaf edge] {
          node[leafF] {$H$ \textit{fully\\ informative}}
        }
      }
    }
    child[xlevel3] {
      node[internal] {$n$ odd}
      child[xlevel4] {
        node[internal] {$q$ odd}
        child[leaf edge] {
         node[leafF] {$H$ \textit{fully\\ informative}}
        }
      }
      child[xlevel4] {
        node[internal] {$q$ even}
        child[leaf edge] {
          node[leafP] {$H$ \textit{partially\\ informative}}
        }
      }
    }
  }
  child[xlevel2] {
    node[internal] {$|C|<n$}
    child[xlevel3] {
  node[internal] {$C$ MISSING-PAIR $\Rightarrow$\\ $\Rightarrow B$ FULL-PAIR }
      child[xlevel4] {
        node[internal] {}
        child[leaf edge] {
         node[leafU] {$H$ \textit{completely\\ uninformative}}
        }
      }
    }
  }
};

\end{tikzpicture}
\caption{Schema of the informativeness criteria for $H=\{A\}\cup C$ with $C={\mathcal R}_n\setminus B$, with $n$ total number of signal qubits, and $q\leq n$ number of signal qubits in $C$.}
\end{figure}
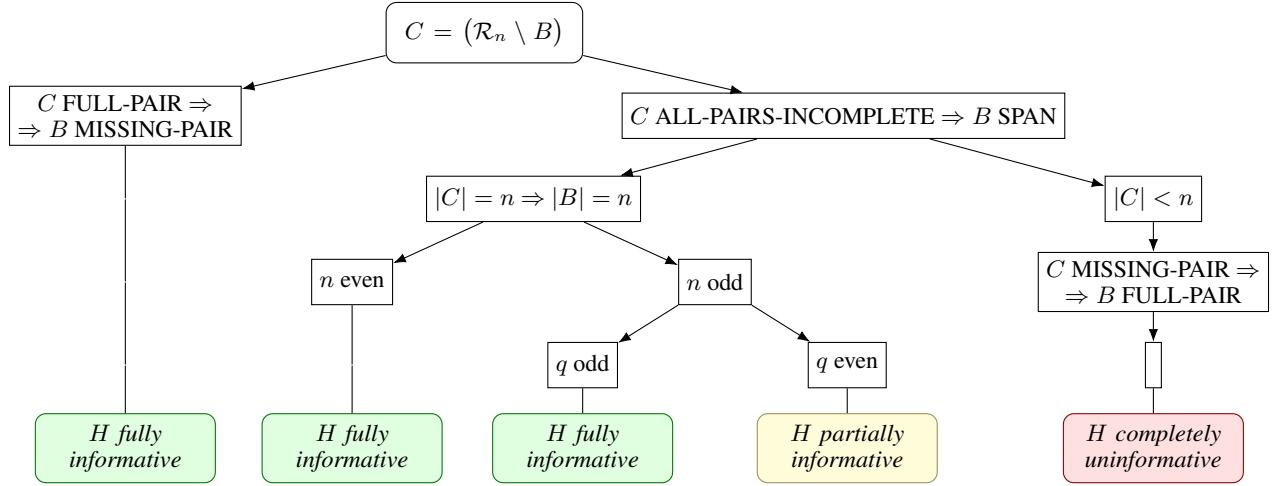

Obviously, for the fully informative state to be extracted one has to devise a specific unitary decoding map.

Thus, sets of qubits containing $A$ have properties that are complementary to those of the sets that do not contain $A$. 
\begin{itemize}
\item
Sets without the qubit $A$ are fully informative only 
when the SPAN and FULL-PAIR conditions are both satisfied, and hide the information either partially,
when the SPAN condition holds, \(|B|=n\) is odd, and the number of signal qubits is odd, or completely, in all the other cases.
\item
On the contrary, sets including the qubit $A$ are fully informative, except for some subcases of the condition ALL-PAIR-INCOMPLETE: when $|C|<n$, the information is completely protected; when
$|C|=n$ odd and the number $q$ of signal qubits in $C$ is even, they are partially informative.
\end{itemize}

We now focus on the states where $|C|=n$ to identify the form of the partially informative sets. 

\subsection{Reduced state for sets $H=\{A\}\cup C$, with $|C|=n$}

To work out the structure of the reduced states of the set $H=\{A\}\cup C$, with $|C|=n$, where $C$ contains $q$ signal qubits and $n-q$ noise qubits (so that the complementary subset $B$ contains $p=n-q$ signal qubits), we start from equation (\ref{eq:encoded-state}) 
and trace out the \(q\) noise qubits \(N_1,\dots,N_q\) and the \(n-q\) signal qubits \(S_{q+1},\dots,S_n\), obtaining the following equation
\begin{equation}\label{eq:with-A}
\rho_{A, S_1, \ldots, S_{q}, N_{q+1}, \ldots, N_n}=\frac{1}{2^{n+2}} \sum_{\mu, \nu=0}^3 \alpha_\mu^{-1} \alpha_\nu\Big(\sigma_\mu|\psi\rangle\langle\psi| \sigma_\nu\Big)_A 
\otimes
\left(\sigma_\mu \sigma_\nu\right)^{\otimes q}\otimes
\left(\left(\sigma_\nu \sigma_\mu\right)^{\top}\right)^{\otimes(n-q)}  .
\end{equation}

This expression derives from the fact that 
tracing out one noise qubit and tracing out one signal qubit yield, respectively
\begin{equation}
\operatorname{Tr}_N\!\Big(\ket{\phi_\mu}\!\bra{\phi_\nu}\Big)
=
\frac{1}{2}\sigma_\mu\sigma_\nu,
\qquad \text{and}\qquad
\operatorname{Tr}_S\!\Big(\ket{\phi_\mu}\!\bra{\phi_\nu}\Big)
=
\frac{1}{2}(\sigma_\nu\sigma_\mu)^{\top}.
\label{eq:bell-identities-corrected}
\end{equation}

We can expand  $\ket{\psi}\bra{\psi}=\frac12\left(I+xX+yY+zZ\right)$,
where the Bloch vector components are defined as
\begin{equation}\label{eq:Bloch-components}
x=\langle\psi| X|\psi\rangle, \quad y=\langle\psi| Y|\psi\rangle, \quad z=\langle\psi| Z|\psi\rangle,
\end{equation}
inside the first operatorial factor in expression (\ref{eq:with-A}), to get
\begin{equation}\label{eq:factor-A}
\Big(\sigma_\mu|\psi\rangle\langle\psi| \sigma_\nu\Big)=\frac{1}{2}\left(\sigma_\mu \sigma_\nu+x \sigma_\mu X \sigma_\nu+y \sigma_\mu Y \sigma_\nu+z \sigma_\mu Z \sigma_\nu\right)=\frac{1}{2}\sum_r b_r\,( \sigma_{\mu}\sigma_{r}\sigma_{\nu})\,,
\end{equation}
where $\left(b_0, b_1, b_2, b_3\right)=(1, x, y, z)$, and the index $r$ denotes the Bloch component. 
This makes it possible to expand the reduced-state expression (\ref{eq:with-A}) according to four contributions, each associated to a Bloch vector component, and to ask which terms are zero due to cancellations. Rewriting (\ref{eq:with-A}) along these guidelines, i.e. gathering by Bloch component, we get
$$
\rho_{A, S_1, \ldots, S_{q}, N_{q+1}, \ldots, N_n}=\frac{1}{2^{n+3}} \sum_{r=0}^3 b_r\sum_{\mu, \nu=0}^3  \alpha_\mu^{-1} \alpha_\nu\left(\sigma_\mu \sigma_r \sigma_\nu\right)_A 
\otimes\left(\sigma_\mu \sigma_\nu\right)^{\otimes q}
\otimes\left(\left(\sigma_\nu \sigma_\mu\right)^{\top}\right)^{\otimes(n-q)}.
$$

At this point we want to identify the possible cancellations among the $16$ terms of the sum $\sum_{\mu,\nu=0}^3$. To perform the bookkeeping, we define separately the coefficient matrices associated to each factor.

This allows us to define the following operator-valued matrices, obtained from the Pauli operators and from the coefficient matrices ${\mathsf S} $, ${\mathsf N}$ and ${\mathsf C} $. 

The matrix
\(
{\mathsf S} \equiv (\sigma_\mu \sigma_\nu),
\)
(where the S stands for "signal", since we traced out the noise qubit) admits the following decomposition.
\begin{eqnarray}
{\mathsf S} \equiv (\sigma_\mu \sigma_\nu)
&=&
I I_4
+X\,{\mathsf S}_{1}
+Y\,{\mathsf S}_{2}
+Z\,{\mathsf S}_{3}.
\end{eqnarray}
where \(I_4\) is the \(4\times4\) identity and \({\mathsf S}_1,{\mathsf S}_2,{\mathsf S}_3\) are coefficient matrices.
\begin{equation}\label{eq:matrix-S}
{\mathsf S}_1\equiv
\begin{pmatrix}
\cdot & 1 & \cdot & \cdot \\
1 & \cdot & \cdot & \cdot \\
\cdot & \cdot & \cdot & i \\
\cdot & \cdot & -i & \cdot
\end{pmatrix}
\qquad {\mathsf S}_2\equiv
\begin{pmatrix}
\cdot & \cdot & 1 & \cdot \\
\cdot & \cdot & \cdot & -i \\
1 & \cdot & \cdot & \cdot \\
\cdot & i & \cdot & \cdot
\end{pmatrix}
\qquad {\mathsf S}_3\equiv
\begin{pmatrix}
\cdot & \cdot & \cdot & 1 \\
\cdot & \cdot & i & \cdot \\
\cdot & -i & \cdot & \cdot \\
1 & \cdot & \cdot & \cdot
\end{pmatrix}.
\end{equation}
The matrix
\(
{\mathsf N} \equiv \big((\sigma_{\nu}\sigma_{\mu})^{\top}\big)_{\mu,\nu=0}^3
\) (where the N stands for "noise", since we traced out the signal qubit) admits the following decomposition.
\begin{eqnarray}
(\sigma_\nu \sigma_\mu)^{\top}
&=&
I I_4
+X\,{\mathsf N}_{1}
+Y\,{\mathsf N}_{2}
+Z\,{\mathsf N}_{3}.
\end{eqnarray}
with
\begin{equation}\label{eq:matrix-N}
{\mathsf N}_1\equiv
\begin{pmatrix}
\cdot & 1 & \cdot & \cdot \\
1 & \cdot & \cdot & \cdot \\
\cdot & \cdot & \cdot & -i \\
\cdot & \cdot & i & \cdot
\end{pmatrix}
\qquad {\mathsf N}_2\equiv
\begin{pmatrix}
\cdot & \cdot & -1 & \cdot \\
\cdot & \cdot & \cdot & -i \\
-1 & \cdot & \cdot & \cdot \\
\cdot & i & \cdot & \cdot
\end{pmatrix}
\qquad {\mathsf N}_3\equiv
\begin{pmatrix}
\cdot & \cdot & \cdot & 1 \\
\cdot & \cdot & -i & \cdot \\
\cdot & i & \cdot & \cdot \\
1 & \cdot & \cdot & \cdot
\end{pmatrix}.
\end{equation}

As to the factor $\alpha_\mu^{-1}\alpha_\nu$, from the expressions (\ref{eq:alphas})  for $\alpha_\mu$'s, we have
\[
\alpha_0^{-1}=1,\qquad
\alpha_1^{-1}=\alpha_3^{-1}=-i,\qquad
\alpha_2^{-1}=-i^{-(n+1)}=-(-i)^{n+1}.
\]
Hence, the $n$-dependent matrix
\[
\begin{array}{c|cccc}
\alpha_\mu^{-1} \alpha_\nu & \nu=0 & \nu=1 & \nu=2 & \nu=3 \\
\hline
\mu=0 & 1 & i & -i^{n+1} & i \\
\mu=1 & -i & 1 & i^{n+2} & 1 \\
\mu=2 & -i^{-(n+1)} & -i^{-n} & 1 & -i^{-n} \\
\mu=3 & -i & 1 & i^{n+2} & 1
\end{array}
\]

that we decompose as follows 
\[
\begin{pmatrix}
1 & i & -i^{\,n+1} & i\\
-i & 1 & i^{\,n+2} & 1\\
-i^{-(n+1)} & -i^{-n} & 1 & -i^{-n}\\
-i & 1 & i^{\,n+2} & 1
\end{pmatrix}
= I_4+{\mathsf C}^{(n)}_1+{\mathsf C}^{(n)}_2+{\mathsf C}^{(n)}_3,
\]
with $I_4$ the identity, and
\[
{\mathsf C}^{(n)}_1
\equiv
\begin{pmatrix}
\cdot & i & \cdot & \cdot\\
-i & \cdot & \cdot & \cdot\\
\cdot & \cdot & \cdot & -i^{-n}\\
\cdot & \cdot & i^{\,n+2} & \cdot
\end{pmatrix}
\quad
{\mathsf C}^{(n)}_2\equiv
\begin{pmatrix}
\cdot & \cdot & -i^{\,n+1} & \cdot\\
\cdot & \cdot & \cdot & 1\\
-i^{-(n+1)} & \cdot & \cdot & \cdot\\
\cdot & 1 & \cdot & \cdot
\end{pmatrix}
\quad
{\mathsf C}^{(n)}_3
\equiv
\begin{pmatrix}
\cdot & \cdot & \cdot & i\\
\cdot & \cdot & i^{\,n+2} & \cdot\\
\cdot & -i^{-n} & \cdot & \cdot\\
-i & \cdot & \cdot & \cdot
\end{pmatrix}.
\]

At this point, we define
\[
{\mathsf L}_j^{(n,q)}
\equiv
{\mathsf C}_j^{(n)}
\circ
{\mathsf S}_j^{\circ q}
\circ
{\mathsf N}_j^{\circ (n-q)},
\qquad j=1,2,3,
\]
where \(\circ\) denotes the Hadamard (entry-wise) product. We also define the Bloch-component operators
\[
\Gamma_{j,r}^{(n,q)}
\equiv
\sum_{\mu,\nu=0}^3
\left({\mathsf L}_j^{(n,q)}\right)_{\mu\nu}\,
\sigma_\mu \sigma_r \sigma_\nu,
\qquad
j=1,2,3,\quad r=0,1,2,3.
\]
Then the reduced state can be rewritten as
\[
\rho_{A, S_1, \ldots, S_{q}, N_{q+1}, \ldots, N_n}
=
\frac{1}{2^{n+1}}\,I_A\otimes I^{\otimes n}
+
\frac{1}{2^{n+3}}
\sum_{j=1}^3
\Bigl(
\Gamma_{j,0}^{(n,q)}
+
x\,\Gamma_{j,1}^{(n,q)}
+
y\,\Gamma_{j,2}^{(n,q)}
+
z\,\Gamma_{j,3}^{(n,q)}
\Bigr)\otimes \sigma_j^{\otimes n}.
\]

Since for each fixed \(j\) the matrices
\(
{\mathsf C}_j^{(n)},
{\mathsf S}_j,
{\mathsf N}_j
\)
have the same support, the matrices
\(
{\mathsf L}_j^{(n,q)}
\)
have only four nonzero entries:\ this introduces considerable simplifications.

For \(j=1\), one finds
\begin{small}
\[
{\mathsf L}_1^{(n,q)}
=
\begin{pmatrix}
\cdot & i & \cdot & \cdot\\
-i & \cdot & \cdot & \cdot\\
\cdot & \cdot & \cdot & -i^{-n}\\
\cdot & \cdot & i^{\,n+2} & \cdot
\end{pmatrix}
\circ
\begin{pmatrix}
\cdot & 1 & \cdot & \cdot\\
1 & \cdot & \cdot & \cdot\\
\cdot & \cdot & \cdot & i^{\,q}\\
\cdot & \cdot & (-i)^{\,q} & \cdot
\end{pmatrix}
\circ
\begin{pmatrix}
\cdot & 1 & \cdot & \cdot\\
1 & \cdot & \cdot & \cdot\\
\cdot & \cdot & \cdot & (-i)^{\,n-q} \\
\cdot & \cdot & i^{\,n-q} & \cdot
\end{pmatrix}
=
\begin{pmatrix}
\cdot & i & \cdot & \cdot\\
-i & \cdot & \cdot & \cdot\\
\cdot & \cdot & \cdot & (-1)^{n-q+1}\\
\cdot & \cdot & (-1)^{n-q+1} & \cdot
\end{pmatrix}
\]
\end{small}%
which depends on the parity of \((n-q)\).

For \(j=2\), we have
\begin{small}
\[
{\mathsf L}_2^{(n,q)}
=
\begin{pmatrix}
\cdot & \cdot & -i^{\,n+1} & \cdot\\
\cdot & \cdot & \cdot & 1\\
-i^{-(n+1)} & \cdot & \cdot & \cdot\\
\cdot & 1 & \cdot & \cdot
\end{pmatrix}
\circ
\begin{pmatrix}
\cdot & \cdot & 1 & \cdot\\
\cdot & \cdot & \cdot & (-i)^{\,q}\\
1 & \cdot & \cdot & \cdot\\
\cdot & i^{\,q} & \cdot & \cdot
\end{pmatrix}
\circ
\begin{pmatrix}
\cdot & \cdot & (-1)^{\,n-q} & \cdot\\
\cdot & \cdot & \cdot & (-i)^{\,n-q}\\
(-1)^{\,n-q} & \cdot & \cdot & \cdot\\
\cdot & i^{\,n-q} & \cdot & \cdot
\end{pmatrix}
\]
\[
=
\begin{pmatrix}
\cdot & \cdot & -(-1)^{\,n-q} i^{\,n+1} & \cdot\\
\cdot & \cdot & \cdot & (-i)^{\,n}\\
-(-1)^{\,n-q} i^{-(n+1)} & \cdot & \cdot & \cdot\\
\cdot & i^{\,n} & \cdot & \cdot
\end{pmatrix}.
\]
\end{small}%
%
For \(j=3\), we have that the result depends only on the parity of \(q\):
\begin{small}
\[
{\mathsf L}_3^{(n,q)}
\!\!=\!\!
\begin{pmatrix}
\cdot & \cdot & \cdot & i\\
\cdot & \cdot & i^{\,n+2} & \cdot\\
\cdot & -i^{-n} & \cdot & \cdot\\
-i & \cdot & \cdot & \cdot
\end{pmatrix}
\circ
\begin{pmatrix}
\cdot & \cdot & \cdot & 1\\
\cdot & \cdot & i^{\,q} & \cdot\\
\cdot & (-i)^{\,q} & \cdot & \cdot\\
1 & \cdot & \cdot & \cdot
\end{pmatrix}
\circ
\begin{pmatrix}
\cdot & \cdot & \cdot & 1\\
\cdot & \cdot & (-i)^{\,n-q} & \cdot\\
\cdot & i^{\,n-q} & \cdot & \cdot\\
1 & \cdot & \cdot & \cdot
\end{pmatrix}
\!\!=\!\!
\begin{pmatrix}
\cdot & \cdot & \cdot & i\\
\cdot & \cdot & (-1)^{\,q+1} & \cdot\\
\cdot & (-1)^{\,q+1} & \cdot & \cdot\\
-i & \cdot & \cdot & \cdot
\end{pmatrix}.
\]
\end{small}
The corresponding Pauli sums simplify drastically. In each sector \(j\), only one of the four operators
$\Gamma_{j,0}^{(n,q)}$,
$\Gamma_{j,1}^{(n,q)}$,
$\Gamma_{j,2}^{(n,q)}$,
$\Gamma_{j,3}^{(n,q)}$
is nonzero.

For \(j=1\), the only nonzero contribution depends on the parity of \((n-q)\):
\[
\begin{cases}
\Gamma_{1,3}^{(n,q)}=-4Y, & (n-q) \text{ even},\\[1mm]
\Gamma_{1,2}^{(n,q)}=+4Z, & (n-q) \text{ odd},
\end{cases}
\]
while all the remaining \(\Gamma_{1,r}^{(n,q)}\) vanish.

For \(j=2\), one obtains
\[
\begin{array}{c|c}
(n,q) & \text{non-zero operator}\\
\hline
n \text{ even},\ q \text{ even} & \Gamma_{2,1}^{(n,q)}=4(-1)^{n/2}Z\\
n \text{ even},\ q \text{ odd} & \Gamma_{2,3}^{(n,q)}=4(-1)^{n/2}X\\
n \text{ odd},\ q \text{ even} & \Gamma_{2,0}^{(n,q)}=4(-1)^{(n+1)/2}Y\\
n \text{ odd},\ q \text{ odd} & \Gamma_{2,2}^{(n,q)}=4(-1)^{(n-1)/2}I
\end{array}
\]
and all the remaining \(\Gamma_{2,r}^{(n,q)}\) vanish.

For \(j=3\), one finds
\[
\Gamma_{3,2}^{(n,q)}=-4X,
\qquad
\Gamma_{3,0}^{(n,q)}=
\Gamma_{3,1}^{(n,q)}=
\Gamma_{3,3}^{(n,q)}=0,
\qquad
q \text{ even},
\]
and
\[
\Gamma_{3,1}^{(n,q)}=+4Y,
\qquad
\Gamma_{3,0}^{(n,q)}=
\Gamma_{3,2}^{(n,q)}=
\Gamma_{3,3}^{(n,q)}=0,
\qquad
q \text{ odd}.
\]

Substituting these expressions into the reduced-state formula yields the following parity-dependent forms.

\paragraph{Case \(n\) even, \(q\) even.}
\[
\rho_{A, S_1, \ldots, S_{q}, N_{q+1}, \ldots, N_n}
=
\frac{1}{2^{n+1}}
\left(
I
-
z\,Y_A \otimes X^{\otimes n}
+
(-1)^{n/2}\,x\,Z_A \otimes Y^{\otimes n}
-
y\,X_A \otimes Z^{\otimes n}
\right).
\]

\paragraph{Case \(n\) even, \(q\) odd.}
\[
\rho_{A, S_1, \ldots, S_{q}, N_{q+1}, \ldots, N_n}
=
\frac{1}{2^{n+1}}
\left(
I
+
y\,Z_A \otimes X^{\otimes n}
+
(-1)^{n/2}\,z\,X_A \otimes Y^{\otimes n}
+
x\,Y_A \otimes Z^{\otimes n}
\right).
\]

\paragraph{Case \(n\) odd, \(q\) odd.}
\[
\rho_{A, S_1, \ldots, S_{q}, N_{q+1}, \ldots, N_n}
=
\frac{1}{2^{n+1}}
\left(
I
-
z\,Y_A \otimes X^{\otimes n}
+
(-1)^{(n-1)/2}\,y\,I_A \otimes Y^{\otimes n}
+
x\,Y_A \otimes Z^{\otimes n}
\right).
\]

\paragraph{Case \(n\) odd, \(q\) even.}
\[
\rho_{A, S_1, \ldots, S_{q}, N_{q+1}, \ldots, N_n}
=
\frac{1}{2^{n+1}}
\left(
I
+
y\,Z_A \otimes X^{\otimes n}
+
(-1)^{(n+1)/2}\,Y_A \otimes Y^{\otimes n}
-
y\,X_A \otimes Z^{\otimes n}
\right).
\]

These outcomes are consistent with the classification obtained by complementarity in Subsection~\ref{sec:withA-class}. For all cases with \(n\) even, and for the cases with \(n\) odd and \(q\) odd, all three Bloch components appear in the reduced state, as must be the case for a fully informative subset. Note that the presence of all Bloch components is necessary but not sufficient for perfect recovery, so this is a consistency check rather than an independent derivation; the derivation is the one based on Lemma~\ref{lem:decoupling}.
When \(n\) is odd and \(q\) is even, we have partial informativeness:
all dependence on the input state is confined to the \(y\)-component of the
Bloch vector.

\subsubsection{Small-\(n\) illustrative examples}

For the sake of illustration, we explicitly list here the reduced states for some small values of \(n\) and \(q\). These examples provide a direct check of the general parity-dependent formulas and make transparent the distinction between fully informative and partially informative cases. The case $n=1$ is reported for completeness, although the protocol is formulated for $n>1$: as already observed in \cite{gianini2026encrypted}, for a single pair the signal qubit is not fully encrypted, so that this case should be regarded as a degenerate limit rather than as an instance of the protocol.

\paragraph{Example \(n=1\).}

For \(q=0\), the accessible set is \(H=\{A,N_1\}\), and one obtains
\[
\rho_{A,N_1}
=
\frac14
\left(
I
+
y\,Z_A\otimes X
-
Y_A\otimes Y
-
y\,X_A\otimes Z
\right).
\]
This is the case \(n\) odd, \(q\) even, and is partially informative, since the dependence on the input state is only through the \(y\)-Bloch component.

For \(q=1\), the accessible set is \(H=\{A,S_1\}\), and one finds
\[
\rho_{A,S_1}
=
\frac14
\left(
I
-
z\,Y_A\otimes X
+
y\,I_A\otimes Y
+
x\,Y_A\otimes Z
\right).
\]
This is the case \(n\) odd, \(q\) odd, and is fully informative.

\paragraph{Example \(n=2\).}

For \(q=0\), the accessible set is \(H=\{A,N_1,N_2\}\), and the reduced state is
\[
\rho_{A,N_1,N_2}
=
\frac18
\left(
I
-
z\,Y_A\otimes X^{\otimes2}
-
x\,Z_A\otimes Y^{\otimes2}
-
y\,X_A\otimes Z^{\otimes2}
\right).
\]

For \(q=1\), the accessible set is \(H=\{A,S_1,N_2\}\), and one obtains
\[
\rho_{A,S_1,N_2}
=
\frac18
\left(
I
+
y\,Z_A\otimes X^{\otimes2}
-
z\,X_A\otimes Y^{\otimes2}
+
x\,Y_A\otimes Z^{\otimes2}
\right).
\]

For \(q=2\), the accessible set is \(H=\{A,S_1,S_2\}\), and one finds
\[
\rho_{A,S_1,S_2}
=
\frac18
\left(
I
-
z\,Y_A\otimes X^{\otimes2}
-
x\,Z_A\otimes Y^{\otimes2}
-
y\,X_A\otimes Z^{\otimes2}
\right).
\]

In agreement with the general classification, all the \(n=2\) cases are fully informative.

\paragraph{Example \(n=3\).}

For \(q=0\), the accessible set is \(H=\{A,N_1,N_2,N_3\}\), and one gets
\[
\rho_{A,N_1,N_2,N_3}
=
\frac1{16}
\left(
I
+
y\,Z_A\otimes X^{\otimes3}
+
Y_A\otimes Y^{\otimes3}
-
y\,X_A\otimes Z^{\otimes3}
\right).
\]

For \(q=1\), the accessible set is \(H=\{A,S_1,N_2,N_3\}\), and one obtains
\[
\rho_{A,S_1,N_2,N_3}
=
\frac1{16}
\left(
I
-
z\,Y_A\otimes X^{\otimes3}
-
y\,I_A\otimes Y^{\otimes3}
+
x\,Y_A\otimes Z^{\otimes3}
\right).
\]

For \(q=2\), the accessible set is \(H=\{A,S_1,S_2,N_3\}\), and one finds
\[
\rho_{A,S_1,S_2,N_3}
=
\frac1{16}
\left(
I
+
y\,Z_A\otimes X^{\otimes3}
+
Y_A\otimes Y^{\otimes3}
-
y\,X_A\otimes Z^{\otimes3}
\right).
\]

For \(q=3\), the accessible set is \(H=\{A,S_1,S_2,S_3\}\), and one gets
\[
\rho_{A,S_1,S_2,S_3}
=
\frac1{16}
\left(
I
-
z\,Y_A\otimes X^{\otimes3}
-
y\,I_A\otimes Y^{\otimes3}
+
x\,Y_A\otimes Z^{\otimes3}
\right).
\]

Thus, for \(n=3\), the cases with \(q\) even are partially informative and leak only through the \(y\)-channel, whereas the cases with \(q\) odd are fully informative, again in agreement with the general parity-dependent classification. The sign of the state-independent term $Y_A\otimes Y^{\otimes n}$, positive here, is the $n=3$ instance of the factor $(-1)^{(n+1)/2}$ appearing in the general expression.

%
%
\section{Conclusions}\label{sec:conclusions}

We have completed the characterization of informative subsets in qubit quantum encrypted cloning by extending the analysis from storage-only subsets to subsets that also include the transformed source qubit \(A\). 

For subsets of the storage register alone, full recoverability requires both the SPAN and FULL-PAIR conditions (i.e., at least one qubit from each signal-noise pair and at least a full signal-noise pair), while complete non-informativeness follows from the MISSING-PAIR condition (at least one full signal-noise pair is missing), with a single exceptional partially informative family occurring when \(|B|=n\), \(n\) is odd, and the number of signal qubits in the subset is odd. In that case, the leakage is restricted to the \(y\)-component of the Bloch vector.

For subsets of the form \(H=\{A\}\cup C\), the picture is complementary. Such subsets are fully informative whenever the complementary storage-only subset is completely uninformative, completely uninformative whenever the complementary storage-only subset is authorized, and partially informative exactly when the complementary storage-only subset is partially informative. In particular, when \(|C|=n\), partial informativeness occurs only for odd \(n\) and even \(q\), and again the residual leakage is confined to the \(y\)-Bloch channel. Thus, the inclusion of the transformed source qubit does not remove the parity-based structure already found for storage-only subsets, but reorganizes it in a complementary form.

These results show that confidentiality in encrypted cloning is not governed by an all-or-nothing mechanism. Instead, it is controlled by a sharply constrained access structure, with three distinct classes of subsets: fully informative, completely uninformative, and partially informative. From this perspective, encrypted cloning can be interpreted as a pure-state quantum encoding with secret-sharing-like features, although not as a perfect secret-sharing scheme, since intermediate subsets may retain limited information about the input state.

The present classification establishes which subsets containing \(A\) are fully informative, but it does not by itself provide an explicit decoding unitary for each such subset. For authorized storage-only subsets, the decoder of \cite{yamaguchi2026encrypted} completes, together with the encoding, an effective SWAP of the input state onto a chosen signal qubit. By contrast, when the accessible subset itself contains \(A\), the natural decoding task is different: one may instead refocus the encoded information back onto \(A\), or, depending on the subset, onto a chosen clone qubit. Thus, the corresponding decoding unitary is not expected in general to coincide with the storage-only decoder, even when both subsets are fully informative. The explicit construction of such subset-dependent decoding unitaries is left for future work.

The results also have practical implications. The fact that subsets containing the qubit \(A\) are typically rather informative means that the protection of \(A\) plays a particularly critical role in the overall confidentiality of the protocol. In security terms, leakage resilience is not determined only by the distribution of signal and noise qubits across the storage register, but also by whether the transformed input qubit can be jointly accessed with even moderately structured subsets of the register. This has direct architectural consequences: implementations of encrypted cloning should not treat \(A\) as an ordinary component of the encoded system, but as a specially sensitive element requiring stronger isolation, access control, and physical protection.

More broadly, these observations suggest that the architecture of any practical realization should be designed around asymmetric protection assumptions. If the storage register is spatially distributed, the qubit \(A\) should ideally be kept in a distinct and more strongly secured location, or be subject to stronger hardware and protocol-level safeguards than the other qubits. Otherwise, the generic informativeness of subsets containing \(A\) may substantially weaken the effective confidentiality of the encoded state, even in cases where storage-only subsets would remain non-authorized or completely uninformative.

Possible future work includes the explicit construction of decoding unitaries for authorized subsets containing \(A\), and a more systematic comparison between encrypted cloning and quantum secret-sharing schemes.


\bibliographystyle{unsrt}
\bibliography{refs}

\end{document}